\documentclass[reprint,aps,pre,superscriptaddress]{revtex4-2}

\usepackage{amssymb}
\usepackage{amsmath}
\usepackage{url}
\usepackage{graphicx}

\begin{document}




\title{Reinforcement Learning Dynamics of Network Vaccination and Hysteresis: A Double-Edged Sword for Addressing Vaccine Hesitancy}

\author{Atticus McWhorter}
\email{atticus.w.mcwhorter.gr@dartmouth.edu}
\affiliation{Department of Mathematics, Dartmouth College}\thanks{Funded by the Dartmouth Fellowship to A.M.}

\author{Feng Fu}
\email{fufeng@gmail.com}
\affiliation{Department of Mathematics, Dartmouth College}
\affiliation{Department of Biomedical Data Science, Geisel School of Medicine}

\date{\today}

\begin{abstract}
Mass vaccination remains a long-lasting challenge for disease control and prevention with upticks in vaccine hesitancy worldwide. Here, we introduce an experience-based learning (Q-learning) dynamics model of vaccination behavior in social networks, where agents choose whether or not to vaccinate given environmental feedbacks from their local neighborhood. We focus on how individuals' bounded rationality impacts decision-making of irrational agents in networks. Additionally, we observe hysteresis behavior and bistability with respect to vaccination cost and Q-learning hyperparameters. Our results offer insight into the complexities of Q-learning and particularly how foresightedness of individuals will help mitigate - or conversely deteriorate, therefore acting as a double-edged sword - collective action problems in important contexts like vaccination. We also find a diversification of uptake choices, with individuals evolving into complete opt-in vs. complete opt-out. Our results have real-world implications for targeting the persistence of vaccine hesitancy using an interdisciplinary computational social science approach that integrates social networks, game theory, and learning dynamics.
\end{abstract}


\maketitle




\section*{Introduction}

Vaccination is one of the most impactful public health interventions in history, offering protection against common infectious diseases such as measles and whooping cough~\cite{anderson1982directly,Sudfeld2010Effectiveness, Zhang2014Acellular}. However, vaccination also faces a persistent challenge in the form of vaccine hesitancy~\cite{Dube2013Vaccine}. This phenomenon, characterized by skepticism or reluctance toward vaccination, poses significant hurdles to achieving widespread immunization coverage~\cite{bloom2014addressing}. From concerns regarding vaccine safety and efficacy to cultural, religious, and sociopolitical factors, vaccine hesitancy manifests itself in diverse forms across populations worldwide~\cite{bloom2014addressing,malik2021Covid,korn2020vaccination}. Addressing vaccine hesitancy requires a comprehensive understanding of the decision making behind vaccine-hesitant populations and the development of targeted strategies to promote confidence in vaccination programs~\cite{jarrett2015strategies,fugenschuh2022overcoming,chang2025niche}.

Mathematical epidemic models of vaccine dynamics are often helpful for finding an optimal level of vaccination to prevent the spread of disease~\cite{anderson1982directly,chauhan2014stability,heesterbeek2015modeling,yusuf2012optimal}. Past efforts have focused on investigating inequities in vaccine programs of rich and poor countries \cite{wagner2021vaccine}, or studying the competition of multiple diseases \cite{karrer2011competing}. However, these models do not explicitly address how to achieve the optimal vaccination level. To achieve this optimal level, researchers must understand the decision-making dynamics behind vaccination and vaccine hesitancy. Taking a game-theoretic approach, previous studies have investigated these dynamics \cite{bauch2004vaccination,galvani2007long,bauch2004vaccination, chen2019imperfect, glaubitz2023population, wu2011imperfect,kabir2019behavioral}. In these models, individuals are engaged in the social dilemma of vaccination, where payoffs are determined by the costs of vaccination and infection.

Many of these prior models are based on the interplay between collective vaccination level and disease spreading~\cite{wang2016statistical}, coupled with a decision-making step. In addition to payoff-based social imitation, individuals can update their vaccination strategy based on a variety of factors, including their history~\cite{saad2024impact}, group behavior~\cite{wang2021effects}, social incentives~\cite{vardavas2007can}, or a combination of the above~\cite{liu2022coevolution}. However, empirical data analyses of vaccine hesitancy reveal a tendency of vaccine-hesitant individuals to group themselves geographically or form echo chambers on social networks~\citep{may2003clustering,omer2008geographic,wang2014nonmedical, lieu2015geographic}. Thus, it is necessary to assume a structured population for behavioral and attitude changes alongside disease transmission. For example, Ref.~\cite{muller2022echo} takes an opinion-dynamic approach, building a network model of opinion that partitions into echo chambers. 

Researchers have also developed network models of contagion spread coupled with more sophisticated decision-making mechanisms. Two common methods of strategy updating are memory-based~\cite{zhang2012rational} and imitation-based~\cite{fu2011imitation}. In memory-based updates, agents look to their past action-payoff pairs to decide whether to vaccinate, and in imitation-based updates, agents use the payoff information of their neighbors to make decisions. Some models combine imitation and memory-based approaches by mixing agents that use imitation to update their strategies with agents who use reinforcement learning (RL). Ref.~\cite{lu2023reinforcement} uses a simple perceptron-like update rule and employs a parameter that governs the influence of memory-based and imitation-based contributions to the loss function. In Ref.~\cite{kan2023double}, the authors place a portion of intelligent agents using deep Q-learning into a network of individuals using simple imitation strategies. Ref.~\cite{shi2019exploring} uses a different approach; agents use RL to decide between a memory-based and imitation-based update rule. In all three of these prior studies, the authors observed parameter dependence that gave rise to a phase transition between two states: from a vaccinating state to a non-vaccinating state. 

While bistability and hysteresis is revealed in previous studies focusing on social imitation of vaccination behavior in the presence of imperfect vaccines~\cite{chen2019imperfect,fu2025social}, it remains largely unclear whether experience-based RL will lead to non-trivial rich dynamical behavior. In our model, we also combine memory and environmental feedback. Individuals are placed in a social network and choose whether or not to vaccinate given the decisions of those in their immediate neighborhood. Then an epidemic passes through the population. By encoding the number of vaccinated neighbors as a state, agents use Q-learning to update their strategies based on their perceived risk and past successes. Interestingly, we find bistability of learned vaccination equilibria, with respect to vaccination cost as well as hyperparameters like discount rate, particularly when agents have a high level of rationality. Our work contributes to the emerging literature on RL-based network epidemiological studies in that we find bistability and hysteresis with path-dependent convergence even in the presence of perfect vaccines. 

\section*{Model and Methods}

In order to incorporate both individual decisions and epidemic dynamics, we develop a multi-stage game. In the first stage, actors decide whether to get vaccinated, incurring the cost of vaccination $-r_v$. In the second stage, an epidemic spreads through the population, and infected individuals incur the cost of infection $-r_i$. This payoff determination process is depicted in Figure 1.

\begin{figure*}





\includegraphics[width=\linewidth]{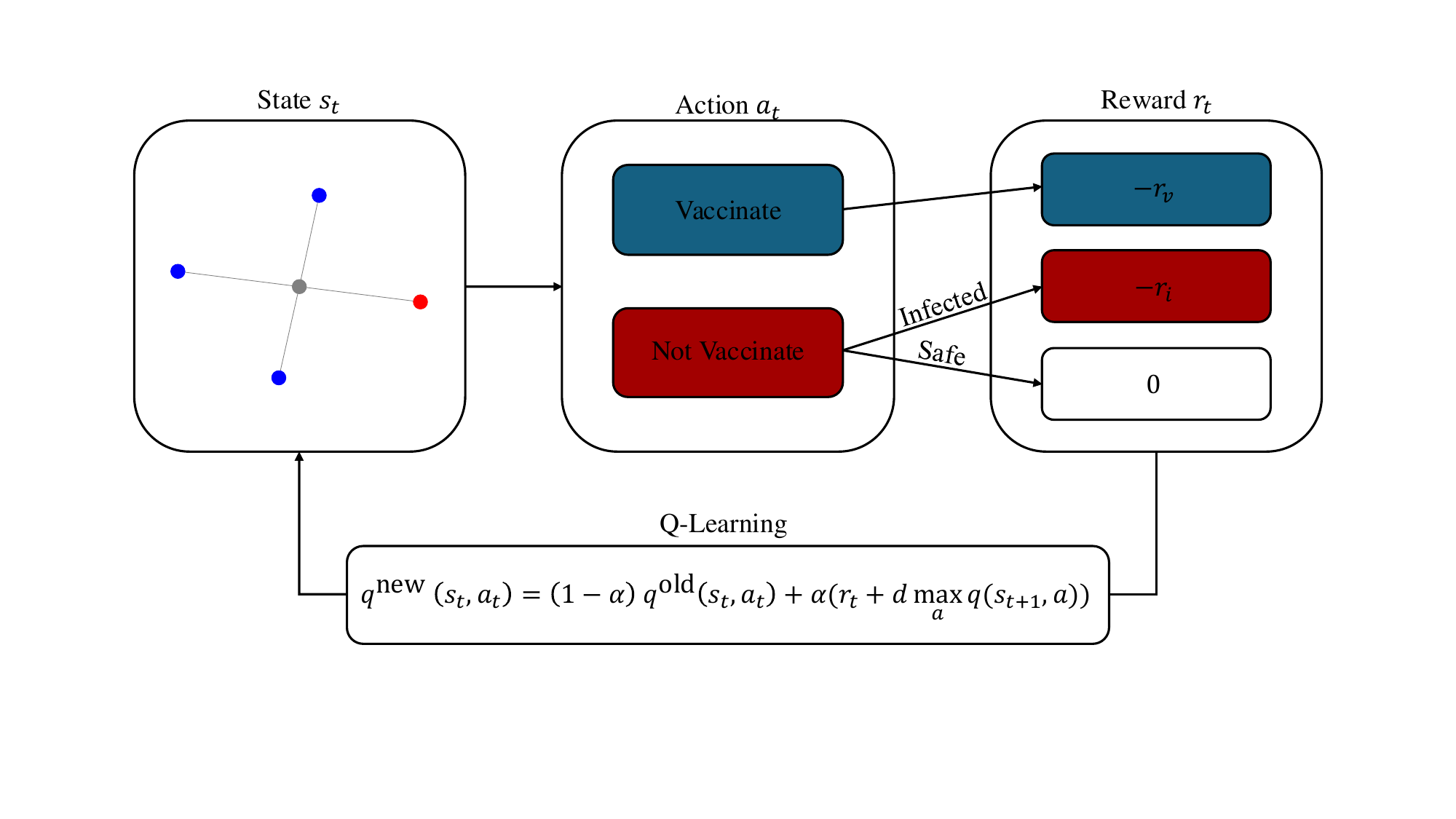}
\caption{\textbf{Payoff structure of the game and reinforcement learning process}. Cooperating (vaccinating) incurs a cost $r_v$, and defecting means the agent does not get vaccinated, in which case they pay a cost $r_i$ if they become infected. We then use Q-learning to allow players to update their strategies between seasonal disease spread.} 
\end{figure*}

    

We have created a model wherein an individual who receives the vaccine has no risk of infection, as in prior work~\cite{fu2011imitation}. Thus, when we implement our epidemic spread on the network, vaccinated individuals will get infected with probability zero. Unvaccinated individuals will become infected at a rate proportional to the number of infected neighbors as follows:\begin{equation}
    \label{eq:prob}
    (1 - (1-\beta)^{f(N(v))})\chi(v \in S).
\end{equation} In Equation \ref{eq:prob}, $\beta$ refers to the transmission rate, $N(v)$ refers to the neighborhood of node $v$, $f(N(v))$ tallies the number of infected individuals in $N(v)$, and $\chi(v \in S)$ returns 1 if node $v$ is susceptible, and 0 otherwise. 

Then, once infected, individuals will recover with probability $\gamma$, and when recovered, we assume that they are not susceptible. Thus, in each time step of the second stage of the model, nodes move from susceptible to infected and/or from infected to recovered. The second stage ends when all nodes are susceptible or recovered, signaling the end of the pandemic. We note that our implementation of the epidemic spreading process is a network-based version of the classic Reed–Frost model, and a similar modeling choice to ours can be found in \cite{karrer2011competing}.

\begin{figure}[ht]
    \centering
    \includegraphics[width=\linewidth]{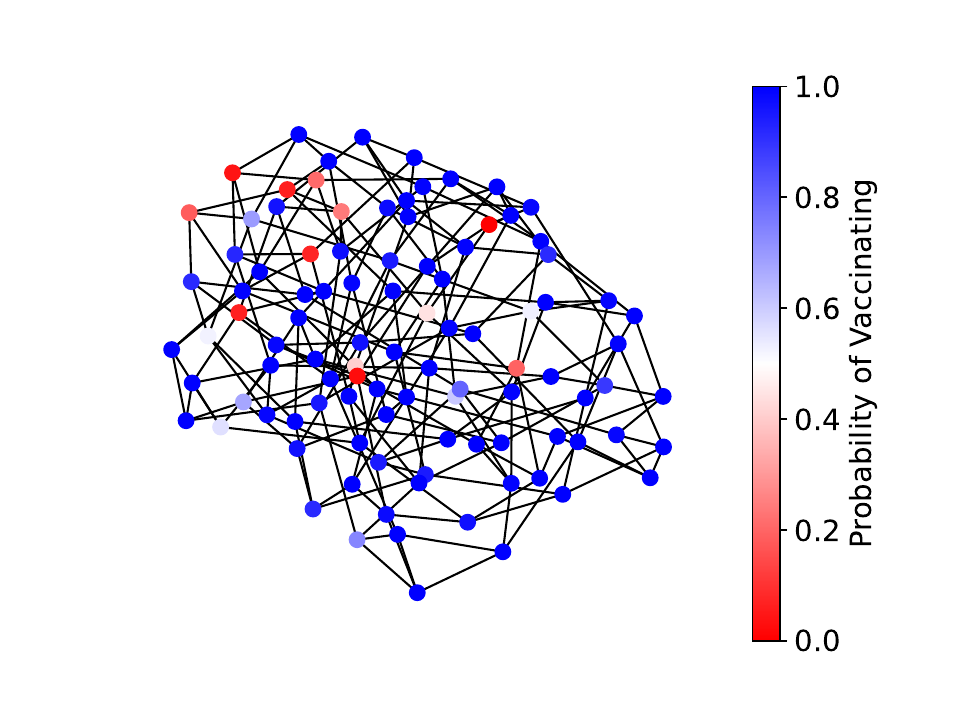}
    \label{fig:process}
    \caption{\textbf{Evolved decision-making network.} The network is instantiated on a $k$-regular graph with degree $k=4$. Here node colors indicate the probability of vaccinating under the state of no vaccinated neighbors $s_t = 0$. A few unvaccinated are intermixed with vaccinated individuals. Parameter values to obtain this network were: $\alpha = 1, \beta = 0.4, \gamma = 0.4, d = 0.95, T = 0.01,  r = 0.8$.
}
\end{figure}

\subsection*{Reinforcement Learning}

Once payoffs are received, agents use Q-learning to choose their action in the next generation (see Figure 1). At the most minimal extreme, the agent could have no information about its environment, and from this single state it would build its two Q-values. On the other end of the spectrum, each agent could know the decisions of everyone in the graph. Clearly, neither are good options. In the minimal case, agents don't have enough information to make good decisions, opting randomly to vaccinate (C) or not (D), and the final outcome could be highly dependent on the initial state. In the maximal case with complete information, it is non-trivial to compute the number of ways to draw a graph with $N$ nodes, and even if one does not take graph structure into account, the state space would grow on the order of $2^N$.

A plausible and reasonable choice is to allow agents to observe their neighbors' choices; this way the state space grows on the order of $N\overline{k}$, where $\overline{k}$ is the average degree. Then, to simplify the Q-tables, we instantiate the model on a $k$-regular graph. This choice ensures that the Q-table for each node will be the same dimension, which eases implementation. In this analysis, we choose $k = 4$, and an example Q-table is shown in Table \ref{tab:Qtable}.

\begin{table}[h]
    \centering
    \begin{tabular}{r|cccccccccc}
         & 0 & 1 & 2& 3 & 4 &\\
         \hline
        C  & $q_{C,0}$ & $q_{C,1}$ & $q_{C,2}$ & $q_{C,3}$ & $q_{C,4}$\\
        D & $q_{D,0}$ & $q_{D,1}$ & $q_{D,2}$ & $q_{D,3}$ & $q_{D,4}$\\
    \end{tabular}
    \caption{\textbf{Q-Values.} State-action pairs on a $k$regular graph with $k = 4$.}
    \label{tab:Qtable}
\end{table}

In order to train the Q-tables for each node, we implemented the Boltzmann choice algorithm, where the agent chooses action $i$ with probability,

\begin{equation}
    P(a_t = i|s) = \frac{1}{Z} e^{\frac{q_{i, s}}{T}},
\end{equation}

\noindent where $Z$ is a normalization term, $T$ is the temperature, and $s$ is the state. After choosing an action, the epidemic stage of the model begins, then each agent receives a payoff. Finally, each agent will update their Q-table according to Equation \ref{eq:Qtable}:

\begin{equation}
\label{eq:Qtable}
    q^{\text{new}}(s_t, a_t) = (1 - \alpha)q^{\text{old}}(s_t, a_t) + \alpha (r_t + d \max_a q(s_{t+1}, a)),
\end{equation}

\noindent where $s_t$ and $a_t$ are the state and action in iteration $t$, respectively, $\alpha$ is the learning rate, $r_t$ is the payoff, and $d$ is a discount factor. 

The model naturally has many different parameters. Firstly, there are societal parameters that can be informed by real-world networks. These societal parameters are the population size $N$ and the degree $k$, which could be fixed using a real-world network. The remaining parameters are the biological contagion parameters, which could be estimated using epidemiological data. These are $\beta$ and $\gamma$, the infection and recovery rates. There is a great body of research that estimates these parameters for real-world diseases~\citep{longini1982estimating, boelle2011transmission, yang2015inference, leventhal2015evolution, ndiaye2020analysis, fu2017dueling}. We have vaccination-related parameters that we do not have to explicitly set. These are $r_v$ and $r_i$, the costs associated with vaccination and infection. But since multiplication by a constant does not change the game, we investigate the single parameter $r = \frac{r_v}{r_i}$, the relative cost of vaccination to infection with $-1 < r < 1$ ($r<0$ allows vaccination to be rewarding, such as through subsidies). Lastly, we have learning parameters, which are $T$, the temperature of the choice algorithm, $d$, the discount factor, and $\alpha$, the learning rate.

\section*{Results}

Under certain choices of learning parameters, the agents successfully learn optimal strategies for the vaccination game. In Figure 3, we have plotted the average timescales of vaccination level and pandemic size. For these epidemic parameters, a vaccination level of $x^* = 1 - \frac{1}{R_0} = \frac{11}{12}$ is the herd immunity threshold. The agents recover this value (with 0.3\% error). This optimum maximizes societal payoff by ensuring that the minimum number of individuals are vaccinated while herd immunity is reached. This is evidenced by the fact that the average pandemic size decays to nearly zero. At this level, the average payoff is higher than $-r_v$, nearly equal to that of the evolutionarily stable strategy ($-r_v * \frac{11}{12}$).

\begin{figure}[ht]
    \centering
    \includegraphics[width=\linewidth]{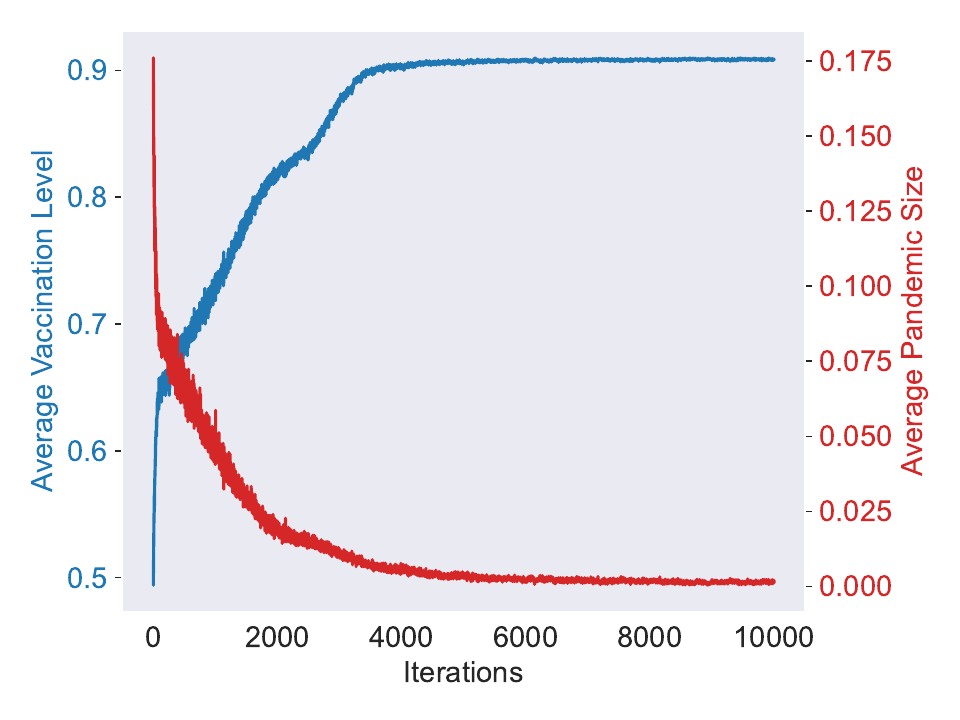}
    \label{fig:timescale}
    \caption{\textbf{Temporal learning dynamics of vaccination and infection.} Over time, agents successfully learn the optimal strategy of an overall vaccination level of 91.3\%. This figure was obtained by averaging the timescales of 100 independent runs. Parameter values to obtain these results: $\alpha = 0.1, \beta = 0.4, \gamma = 0.1, d = 0.8, r = 0.1, T = 0.01.$
}
\end{figure}

\begin{figure}[ht]
  \centering
  \includegraphics[width=\linewidth]{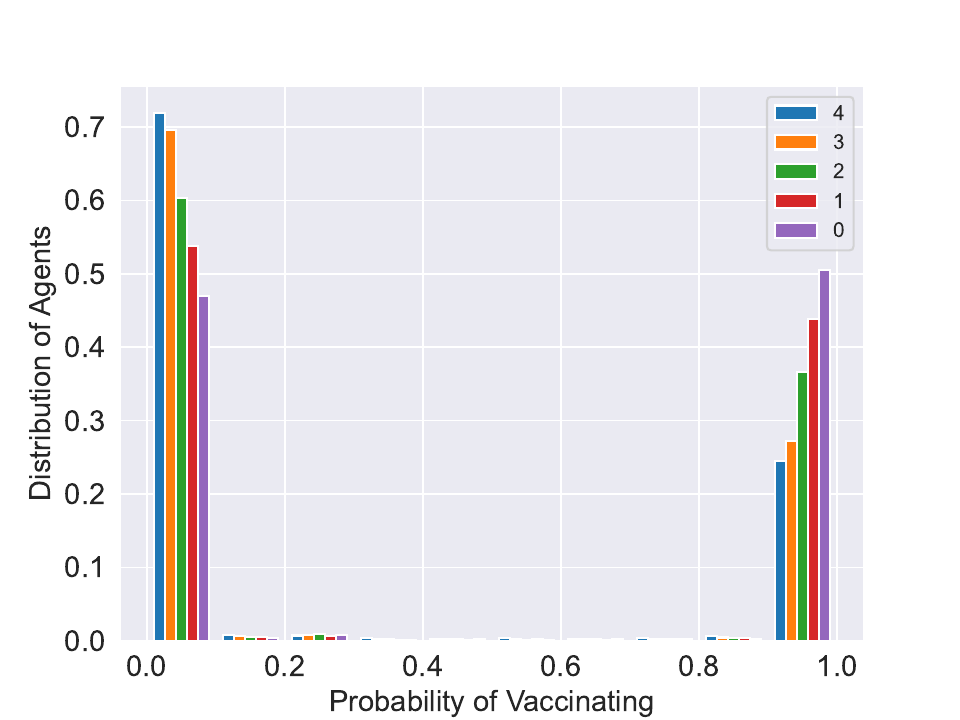}
  \label{fig:decisions}
  \caption{\textbf{Context-dependent diversification of vaccination strategies.} As the state (number of vaccinated neighbors) increases, the height of the right mode (vaccinate) decreases, and the height of the left mode (do not vaccinate) increases. This reveals a double-edged sword impact of network clustering of vaccination. Parameter values to obtain these results: $\alpha = 1, \beta = 0.4, \gamma = 0.5, d = 0.95, r = 0.2, T = 0.01.$
}
\end{figure}

The primary aim of this paper is to investigate the decision-making characteristics of agents given their respective environments. Since we observe that agents find an optimal level of vaccination, self-interested individuals must be using information from their environment to decide when it is safe to not vaccinate. In Figure 4, we have plotted the distribution of agents' strategies. 

Regardless of the number of vaccinated neighbors, each individual tends to choose a pure strategy, vaccinating with probability 0 or 1. This is due to the linearity of the expected payoff function, forcing any maximum to occur on the boundary. Consequently, agents diversify into complete opt-in (pure cooperators) or complete opt-out (free riders), depending upon their assessment of their local environment. Agents respond to their changing environment; as the number of vaccinated neighbors decreases, agents recognize a higher risk of infection, and the number of agents vaccinating with probability one increases. This shows the importance of local risk perception in shaping individual strategies, as the perceived safety of the community can influence decision-making.  

Herein lies the double edged sword; when vaccination rates are high, agents are more likely to have vaccinated neighbors, and are therefore less likely to vaccinate. This creates a negative feedback loop: high levels of cooperation create an environment that could spawn free-riding, and likewise, low levels of vaccination will sway agents to cooperate. 

This distribution is highly dependent on the model parameters. In Figure 5 we plot the dependence of the society-wide vaccination level on the basic reproduction ratio, $R_0$. Estimates of $R_0$ values for influenza \cite{chowell2008seasonal} and measles \cite{Guerra2017basic} are shown. Regardless of temperature, we see that as the basic reproduction ratio increases, so does vaccine uptake. However, we do not see the same societal optimum as previously. Instead of following the $1 - \frac{1}{R_0}$ curve, when the temperature is very low, the curve plateaus around 92\%, and at higher temperatures, the agents are unable to effectively exploit the strategy of vaccination. When $R_0$ is less than one, we do not observe a vaccination level of zero. As the temperature increases, the response curve flattens, as agents cannot consistently exploit the strategies they have found due to the increased randomness in the choice function.

\begin{figure}[ht]
    \centering
    \includegraphics[width=\linewidth]{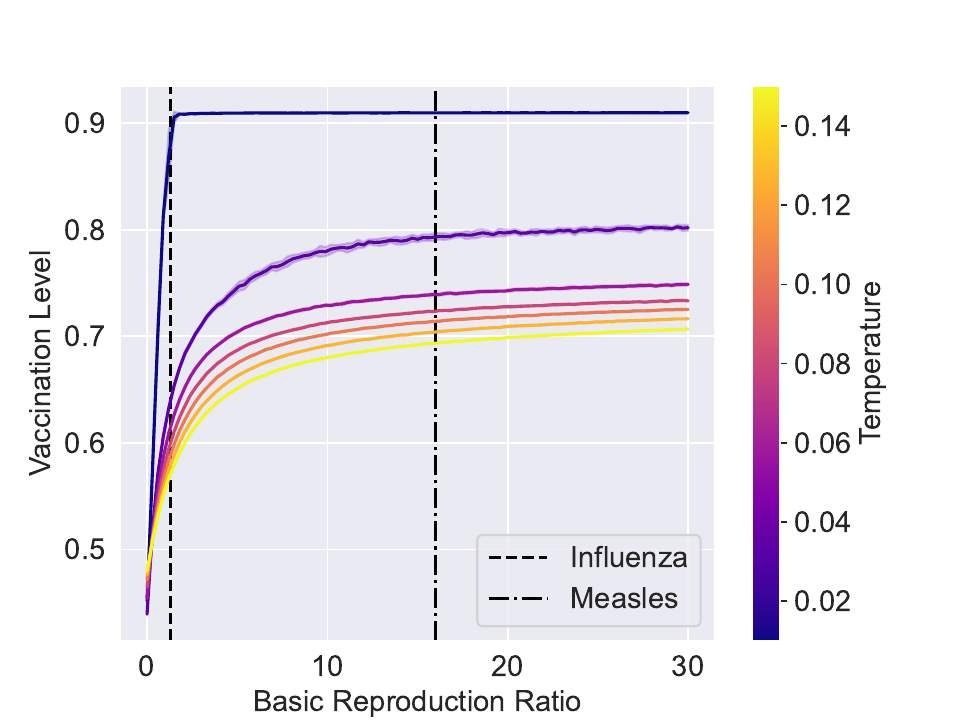}
    \caption{\textbf{Responsiveness to the basic reproduction ratio $R_0$.} As the basic reproduction ratio increases, so does the likelihood of infection for unvaccinated individuals. In response, agents vaccinate at higher rates. At lower temperatures this response is much faster, resulting in higher uptake than at high temperatures. Parameter values to obtain these results: $\alpha = 0.1, d = 0.8, r = 0.1$.}
    \label{fig:enter-label}
\end{figure}

\begin{figure}[ht]
    \centering
    \includegraphics[width=\linewidth]{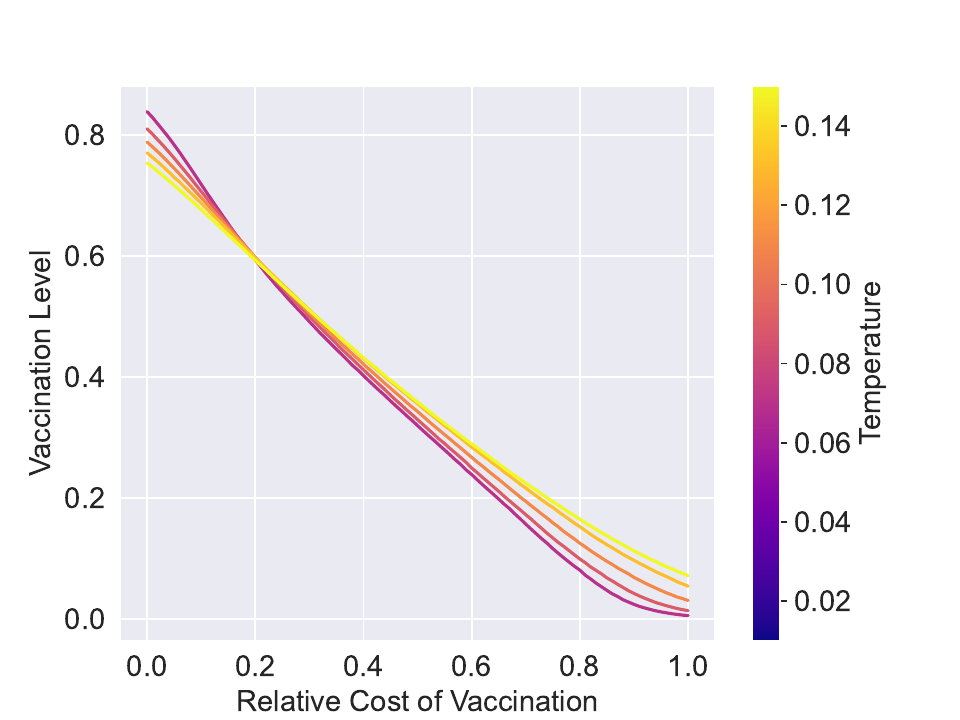}
    \caption{\textbf{Responsiveness to the cost of vaccination at high temperatures.} As the cost of vaccination increases, agents are less likely to vaccinate. Increasing temperature flattens the response curve. Parameter values to obtain these results: $\alpha = 0.1, \beta = 0.4, \gamma = 0.1, d = 0.8$.}
    \label{fig:nice_ones}
\end{figure}

In Figure 6, we illustrate how varying the relative cost of vaccination $r$ affects the vaccination rate at higher temperatures. As the relative cost of vaccination increases, agents tend to vaccinate less. Temperature influences agents' responses similarly to how it affects reactions to changing basic reproduction ratio; lower temperatures enable agents to find safety and respond to high costs, while higher temperatures flatten the response curve towards random behavior.

\begin{figure*}[ht]
    \centering
    \includegraphics[width=\linewidth]{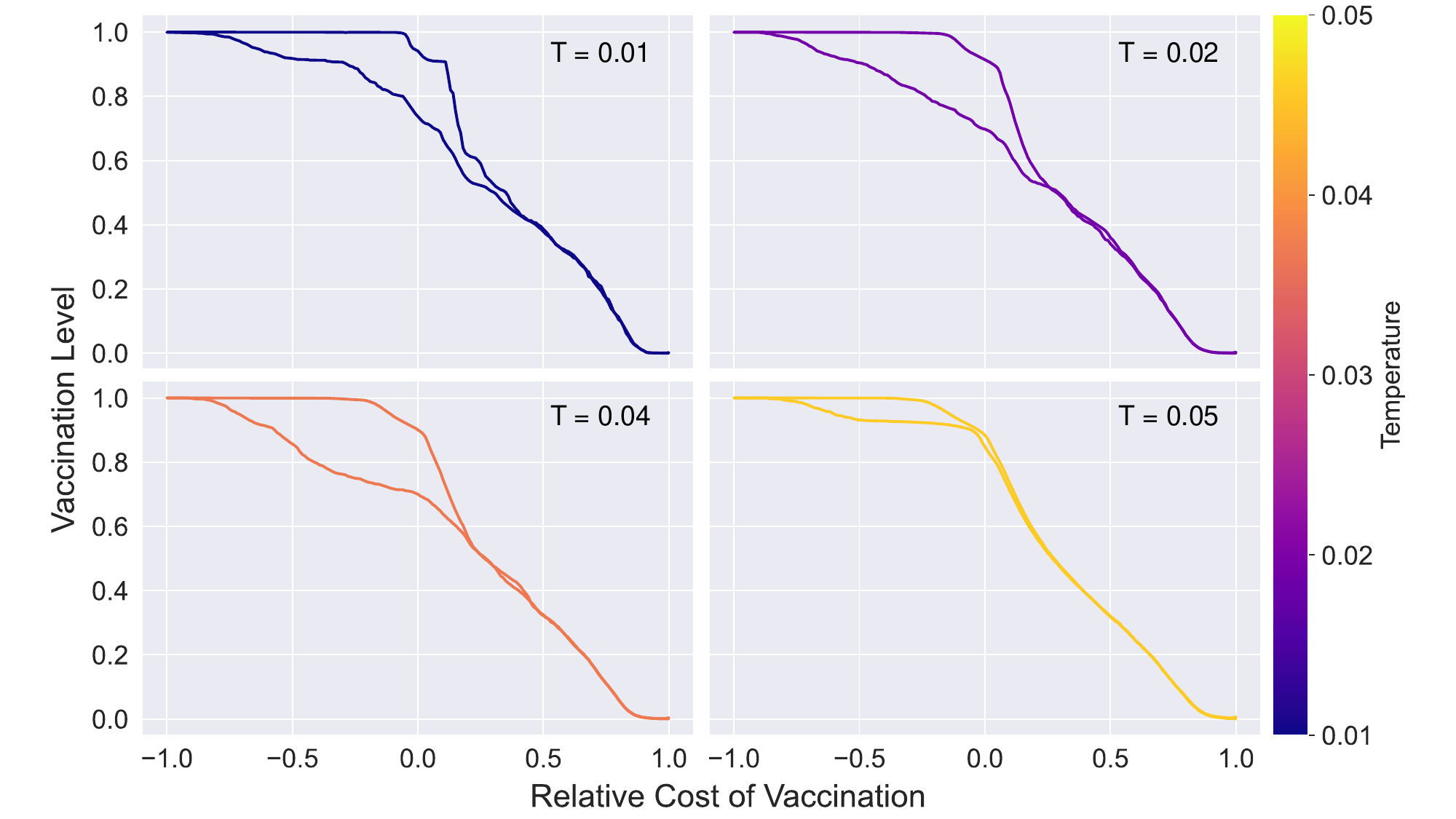}
    \caption{\textbf{Hysteresis loops with respect to varying the cost of vaccination:} When transitioning between equilibrium states, there is a lag that forms a hysteresis loop. The loop expands as temperature increases, then ultimately shrinks and disappears. Parameter values to obtain these results: $\alpha = 0.1, \beta = 0.4, \gamma = 0.1, d = 0.8$.}
    \label{fig:hysteresis}
\end{figure*}

In Figure 7, at very low temperatures, variations in $r$ lead to hysteresis. With a negative $r$, the stable state is 100\% vaccination, as it is the only positive payoff strategy. As $r$ becomes positive, the vaccination rate decreases with increasing $r$, and the rate of decrease depends on temperature. There is a lag during transitions; agents maintain the 100\% vaccination rate longer when costs rise, and hold off on decreasing vaccination rates longer when costs fall, creating a hysteresis loop. The loop size decreases with higher temperatures until it converges with the behavior seen in Figure 6.

This hysteresis loop is a novel observation in multi-agent Q-learning in the specific context of network vaccination. Unlike other models where equilibria shift instantaneously with changing parameters, our method introduces inertia in the system (due to discount rate), delaying transitions between the stable states. Also of note is the loop's dependence on temperature; the phenomenon is only observed at low temperatures, where agents are free to exploit their learned strategies. This result contributes to the understanding of path dependence in multi-agent reinforcement learning, showing how even simple adaptive dynamics can generate complex, non-reversible phase transitions.

The observed hysteresis also shows the importance of timing early intervention in vaccine campaigns. Agents do not always respond instantaneously to fluctuations in costs. Instead, their decisions are also shaped by their memories and experiences of the past. This memory effect allows the population to sustain high vaccination rate as costs rise, but, conversely, memory also delays vaccination and rebound when costs decrease. Thus, this memory effect displays another double-edged sword for addressing vaccine hesitancy.

\section*{Discussion and Conclusion}

In conclusion, we have developed a model of the decision-making behind vaccination with the goal of better understanding vaccine hesitancy in the context of experience-based vaccine uptake choices. Multi-agent Q-learning in a structured population allows us to explore how agents use their local network surroundings to inform their vaccine decisions. 

We found that while agents generally form cooperative strategies to alleviate the risk of infection, defectors can strategically place themselves in groups of cooperators to avoid infection. This contributes to our understanding of how neighborhood safety shapes individual decision-making; an overreliance on cooperative neighbors can erode the overall safety of the network. Thus, our findings suggest that one must carefully consider the subtleties of the learning dynamics that give rise to persistent unvaccination in future interventions. 

In addition to their surroundings, agents also rely on feedback from global factors to inform their decisions. When the basic reproduction ratio increases, agents recognize the increased transmissibility as an increased risk of infection. Accordingly, population-wide vaccine uptake increases. Similarly, when the relative cost of vaccination increases, agents recognize the relative decreases in the cost of infection and are more willing to take the risk of being unvaccinated. This highlights the importance of effective public education on the true risks of vaccination. In most cases, the relative cost of vaccination is very small, but misinformation about adverse vaccine side effects leads some individuals to have a high perception of the cost of vaccination, leading to vaccine hesitancy. 

Our results also display an interesting dependence on temperature. The response curves of $r$ and $R_0$ flatten as temperature increases. This is expected behavior; as temperature increases, agents are unable to exploit the optimal strategies they have found as their choice function becomes increasingly random. A novel behavior we observed is the hysteresis loop in the relative cost of vaccination and its dependence on temperature. At very low temperatures, the response to changes in $r$ lags, forming a hysteresis loop. The size of the loop initially grows as temperature increases, then ultimately shrinks until the loop disappears (Figure 7). This behavior shows that convergence time to optimal strategies in Q-learning can be path dependent. This path-dependent convergence further underscores the importance of proper vaccine-related eduction. If the relative risk of vaccination is initially perceived as high, but then decreases, vaccine uptake will lag behind, as the memory of high relative risk persists. In contrast, if the initial perception of vaccine risk is low, vaccine uptake will remain high even as the perception of risk increases. 

Future directions for this work include the exploration of different network structures, the possibility of an imperfect vaccine~\cite{chen2019imperfect},  the influence of behavioral interventions, and the introduction of heterogeneity among agents. Possibly the least realistic modeling assumption is that of a $k$-regular network. Scale-free networks are widely recognized as better representative of human populations, and implementing this model on a scale-free network or a real social network could provide insight into the importance of highly connected individuals in the vaccination dilemma. An imperfect vaccine would also increase the realism of our model, adding the potential effect that an agent pays both the cost of vaccination and infection in the same iteration, leading to increased vaccine hesitancy. Behavioral interventions such as vaccine campaigns or the implementation of social distancing are common public health initiatives that warrant further study. Lastly, agent heterogeneity would also be a step toward a more realistic model~\cite{wang2010effects}. In reality, different people are affected in different ways by vaccines and infections, which shapes their decision making in a way that we have not accounted for in the present model. 

In sum, we investigate the reinforcement learning dynamics of vaccination behavior in social networks, where agents learn from their past experiences only with partial information from their neighborhood. We discover rich dynamical behavior, particularly the bistability of learned vaccination equilibria and the hysteresis effect with respect to varying the vaccination cost. These results highlight the double-edged impact of reinforcement learning with implications for mitigating the persistence of vaccine hesitancy. 

\section*{Data and Code Availability}
All computations were performed on doob, a 96-core 3.6GHz AMD Epyc system. Code for the project can be found at \url{https://github.com/amcwhorter/vaccine_games}. 

\bibliographystyle{elsarticle-num}
\bibliography{refs_v2}

\appendix
\section{}
\label{app1}

\setcounter{figure}{0}
\makeatletter 
\renewcommand{\thefigure}{A\@arabic\c@figure}
\makeatother
\begin{figure*}[ht]
    \centering
    \includegraphics[width=\linewidth]{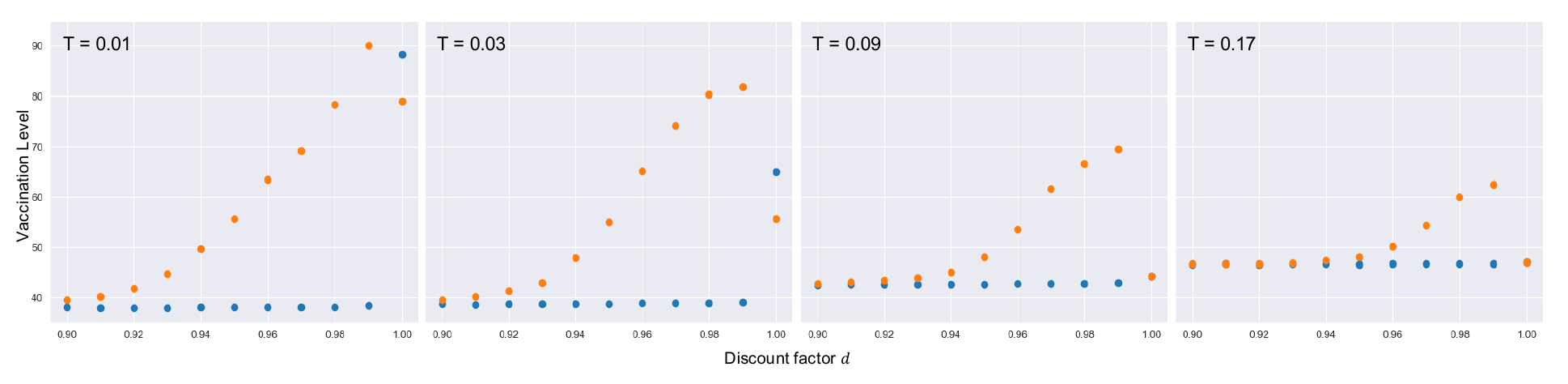}
    \caption{\textbf{Hysteresis with respect to discount rate.} Change in equilibrium state lags behind the change in the discount factor. This forms a hysteresis loop with increasing discount rate (blue) and decreasing discount rate (orange). Increasing temperature decreases this discrepancy Parameter values to obtain these results: $\alpha = 1, \beta = 0.4, \gamma = 0.5, r=0.2$.}
    \label{fig:appendix}
\end{figure*}

Investigation of the full parameter space yielded other interesting results in the interplay of discount rate $d$ and temperature $T$. As discount factor increases (blue) near 1, the vaccination level does not increase, as expected, but remains constant, jumping right when $d = 1$. Going the other way (orange), vaccination level decreases steadily as discount rate decreases. This phenomenon shrinks and disappears as temperature increases. 





\end{document}